\documentclass[reprint,amsmath,amssymb,aps,prl,superscriptaddress,longbibliography]{revtex4-2}

\usepackage{graphicx}
\usepackage{bm}
\usepackage[dvipsnames]{xcolor}
\usepackage[
  colorlinks=true,
  bookmarks=false,
  citecolor=NavyBlue,
  linkcolor=NavyBlue,
  urlcolor=NavyBlue,
  pdftitle={Sign problem and criticality in world-line quantum Monte Carlo methods},
  pdfauthor={Rubem Mondaini and Richard T. Scalettar}
]{hyperref}
\hypersetup{breaklinks=true}
\usepackage{orcidlink}

\newcommand{\prlsection}[1]{\noindent\textit{#1.---}\hspace{0.25em}}

\newcommand{\beginsupplement}{%
  \setcounter{section}{0}%
  \renewcommand{\thesection}{S\arabic{section}}%
  \renewcommand{\theHsection}{S\arabic{section}}%
  \setcounter{equation}{0}%
  \renewcommand{\theequation}{S\arabic{equation}}%
  \renewcommand{\theHequation}{S\arabic{equation}}%
  \setcounter{table}{0}%
  \renewcommand{\thetable}{S\arabic{table}}%
  \renewcommand{\theHtable}{S\arabic{table}}%
  \setcounter{figure}{0}%
  \renewcommand{\thefigure}{S\arabic{figure}}%
  \renewcommand{\theHfigure}{S\arabic{figure}}%
}

\usepackage{color}

\usepackage[normalem]{ulem}

\begin{document}

\title{Sign problem and criticality in world-line quantum Monte Carlo methods}

\author{Rubem Mondaini\,\orcidlink{0000-0001-8005-2297}}
\email{rmondaini@uh.edu}
\affiliation{Department of Physics, University of Houston, Houston, Texas 77004, USA}
\affiliation{Texas Center for Superconductivity, University of Houston, Houston, Texas 77004, USA}

\author{Richard T. Scalettar\,\orcidlink{0000-0002-0521-3692}}
\email{scalettar@physics.ucdavis.edu}
\affiliation{Department of Physics, University of California, Davis, California 95616, USA}

\begin{abstract}
Discussions of the sign problem usually focus on its role as the primary limitation of the applicability of quantum Monte Carlo methods for reliably solving quantum many-body models.  Its behavior as a function of spatial lattice size, doping, temperature, and interaction strengths has been carefully characterized within the determinant quantum Monte Carlo algorithm. Here, we consider the much less well-explored question of the temperature- and size-dependence of the sign problem in the world-line quantum Monte Carlo method. We review how, in this algorithm, even sampling the partition function of free fermions yields a sign problem, which we connect to non-analytic behavior in the corresponding reference model, hard-core bosons on a lattice. The latter exhibits a finite-temperature Kosterlitz-Thouless phase transition in 2D, and we show how this non-analytic behavior is imprinted on the average sign of the weights for the corresponding free fermion 2D tight-binding model. Our results thus show how phase-transition information can remain encoded in the sign problem even after direct sign sampling becomes impractical.
\end{abstract}

\maketitle

\prlsection{Introduction} Quantum Monte Carlo (QMC) methods are among the most powerful nonperturbative tools for studying many-body problems across fields ranging from condensed matter to high-energy physics~\cite{Foulkes2001,Austin2012QMC,Kaul2013QMC,Carlson2015NuclearQMC,Li2019SignProblemFree,Nagata2022}. For many systems and representations, however, the configuration weights entering path-integral or density-matrix formulations are not positive definite. The resulting sign (or phase) problem arises when the weights $W(x)$ become negative (or complex)~\cite{Foulkes2001,Gull2011,Nagata2022}. In severe regimes, typically at low temperatures and large system sizes, cancellations among configurations cause the average sign or phase to decay exponentially with the system volume and inverse temperature~\cite{Loh1990}, leading to an exponential loss of statistical accuracy~\cite{Troyer2005}.

Formally, negative or complex weights can be handled through reweighting. Configurations are sampled according to the positive reference measure $|W(x)|$, while the sign or phase is incorporated into the estimator. Physical expectation values are then obtained from the ratio $\langle \hat{\mathcal{O}} \rangle_{W} =\langle \hat{\mathcal{O}}\mathcal{S}\rangle_{|W|}/\langle \mathcal{S}\rangle_{|W|}$, where $\mathcal{S}(x)\equiv W(x)/|W(x)|$ is the configuration sign or phase factor, and $\langle\cdots\rangle_{|W|}$ denotes an average over the positive measure $|W(x)|$. Although this relation is exact, an exponentially small denominator renders the calculation impractical. Recent work, however, has shown that the average sign, usually regarded only as a measure of computational difficulty, can itself contain useful information about critical behavior and phase transitions~\cite{Mondaini2022,Mou2022,Mondaini2023,SousaJunior2024ExtendedHubbard,Wang2024SignPAM,Yi2024}.

The origin of this connection becomes transparent by expressing the average sign in the reference ensemble as
\begin{equation}
    \langle \mathcal{S}\rangle_{|W|}
    =
    \frac{\sum_x \mathcal{S}(x)|W(x)|}{\sum_x |W(x)|}
    =
    \frac{\mathcal{Z}_{W}}{\mathcal{Z}_{|W|}},
    \label{eq:ave_sign}
\end{equation}
where $\mathcal{Z}_{W}$ and $\mathcal{Z}_{|W|}$ are the partition functions of the physical and reference ensembles, respectively. Equivalently, $\ln \langle \mathcal{S}\rangle_{|W|} = -\beta\left(\mathcal{F}_{W}-\mathcal{F}_{|W|}\right)$, with $\beta=1/T$, the inverse temperature. In the thermodynamic limit, a nonanalytic contribution to the physical free energy $\mathcal{F}_{W}$ is therefore inherited by the logarithm of the average sign, provided that it is not canceled by a corresponding singularity in the reference free energy $\mathcal{F}_{|W|}$. In particular, if $\mathcal{F}_{|W|}$ remains analytic at the physical critical point, the average sign directly encodes the singular free-energy contribution associated with the transition.

The converse situation is equally revealing. If the reference ensemble exhibits a nonanalyticity in $\mathcal{F}_{|W|}$ while the physical ensemble remains analytic, then $\langle\mathcal{S}\rangle_{|W|}$ inherits the singular behavior of the reference system. A natural setting in which this scenario arises is worldline QMC~\cite{Hirsch1981,Hirsch1982,Scalettar1999,AssaadEvertz2008}, where reference ensembles have an easy physical interpretation. Consider, for example, noninteracting spinless fermions on a lattice. In the worldline representation, each configuration acquires a fermionic sign determined by the parity of the permutation realized by the worldlines over the imaginary-time interval. Configurations with odd permutations therefore generate a sign problem. In this formulation, however, the reference ensemble has a transparent physical interpretation: removing the fermionic permutation signs changes the particle statistics and maps the reference system onto hard-core bosons.

Unlike their noninteracting-fermion counterparts, hard-core bosons may undergo a thermal superfluid transition or, depending on the model, a zero-temperature quantum phase transition. The resulting nonanalyticity of the bosonic free energy is then encoded, through Eq.~\eqref{eq:ave_sign}, in the average sign measured with respect to the bosonic reference ensemble. In what follows, we demonstrate this mechanism in both one- and two-dimensional systems exhibiting thermal crossovers or thermal phase transitions.

\prlsection{One dimension} Consider the tight-binding Hamiltonian in an $L$-site ring,
\begin{equation}
\hat {\cal H}_{\rm F} = -t \sum_{\langle i, j\rangle} (\hat c_i^\dagger \hat c_j^{\phantom{\dagger}} + \hat c_j^\dagger \hat c_i^{\phantom{\dagger}})\ ,
\end{equation}
which assigns hoppings with amplitude $t$ between nearest-neighbor sites $\langle i,j\rangle$, creating (annihilating) spinless fermions via $\hat c_i^\dagger$ ($\hat c_i^{\phantom{\dagger}}$) at site $i$. This Hamiltonian is diagonal in momentum, $\hat {\cal H}_{\rm F} = \sum_k \epsilon_k^{\phantom{\dagger}} \hat c_k^\dagger \hat c_k^{\phantom{\dagger}}$, where $\epsilon_k=-2t\cos k$, with $k$ the appropriate allowed momenta. The corresponding partition function in the canonical ensemble, at half-filling $N=L/2$, thus reads $\mathcal Z_{\mathrm{F,BC}}=\sum_{\substack{k_1<\cdots<k_N}}
\exp(-\beta\sum_{\alpha=1}^{N}\epsilon_{k_\alpha})$. Here `BC' refers to the choice of boundary conditions. Although this model is trivially solvable, its worldline representation has a sign problem~\footnote{In one dimension, there is a small but important algorithmic point. With local world-line updates, the spatial winding number cannot change. A simulation started from straight world lines therefore stays in the zero-winding sector, where the fermion sign is positive. Once loop updates are allowed, this is no longer true: a world line can wind around the ring, giving a cyclic permutation and, for even particle number, a negative contribution.}.

Nonetheless, taking the absolute value of the worldline weights removes the fermionic permutation sign while preserving the local hopping amplitudes and the single-occupancy constraint. The reference ensemble is therefore precisely a system of hard-core bosons on the same ring, described by $\hat{\mathcal H}_{\rm B}=-t\sum_{\langle i,j\rangle}(\hat b_i^\dagger\hat b_j^{\phantom{\dagger}}+\hat b_j^\dagger\hat b_i^{\phantom{\dagger}})$, with $(\hat b_i^\dagger)^2=\hat b_i^2=0$, such that $\mathcal Z_{|W|}=\mathcal Z_{\mathrm{B,BC}}$. However, for hard-core particles on a ring, with periodic boundary conditions (PBC), the Jordan-Wigner transformation~\cite{Jordan1928,Lieb1961} makes ${\mathcal Z}_{\mathrm{B,PBC}}$ especially simple. Here, the bosonic reference problem itself maps back to free fermions, but with a boundary condition fixed by the particle-number parity. That is, for the half-filled sizes for which $N=L/2$ is even, the average sign is the ratio of {\it fermionic} partition functions with periodic and anti-periodic boundary conditions (PBC and APBC):
\begin{equation}
    \langle {\cal S}\rangle_{\rm B}=\frac{{\cal Z}_{\mathrm{F,PBC}}}{{\cal Z}_{\mathrm{F,APBC}}}\ ,
    \label{eq:sign-1d-partitions}
\end{equation}
wherein both can be evaluated exactly in the canonical ensemble by computing ${\cal Z}_{\rm F}$ with the corresponding allowed momenta, $k_{\rm PBC}=2\pi n/L$ and $k_{\rm APBC} = (2n+1)\pi/L$, with $n\in[0, L)$.

\begin{figure}[t]
    \centering
    \includegraphics[width=\columnwidth]{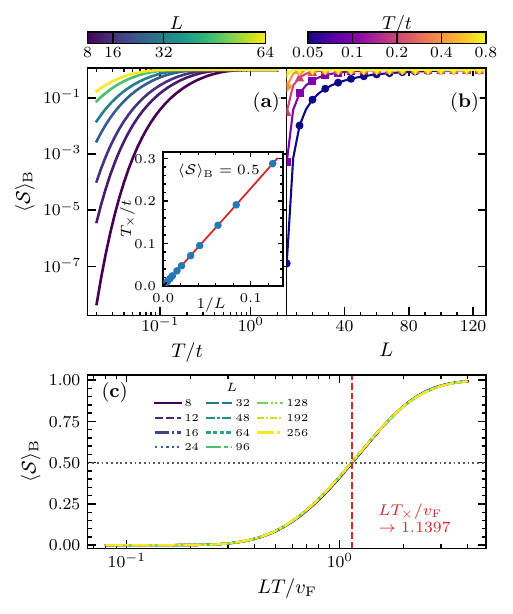}
    \caption{Exact average sign $\langle{\cal S}\rangle_{\rm B}$ for noninteracting spinless fermions on a one-dimensional ring at half filling ($N=L/2$ is even). (a) Temperature dependence for different system sizes. The inset shows the crossover temperature $T_\times$, defined by $\langle{\cal S}\rangle_{\rm B}=1/2$, as a function of $1/L$; its linear extrapolation demonstrates that $T_\times(L)\to0$ in the thermodynamic limit. (b) System-size dependence at fixed temperatures. (c) Collapse of the exact results as a function of the scaling variable $LT/v_{\rm F}$. The horizontal dotted line denotes $\langle{\cal S}\rangle_{\rm B}=1/2$, while the vertical dashed line marks the asymptotic crossover value $LT_\times/v_{\rm F}=1.1397 \simeq \frac{\pi}{2 \ln 4}$. Lines and symbols are exact results without Monte Carlo uncertainty.}
    \label{fig:exact-sign-1d}
\end{figure}

Figure~\ref{fig:exact-sign-1d} reports the temperature and system-size dependence of $\langle {\cal S}\rangle_{\rm B}$. Its central result is that the sign crossover is driven to zero temperature as $L$ increases. At fixed $L$, the APBC sector has the lower ground-state energy, and $\langle{\cal S}\rangle_{\rm B}$ therefore vanishes as $T\to0$. By contrast, at any fixed $T>0$ the free-energy cost of changing the boundary condition vanishes as $L\to\infty$, and the average sign approaches unity. The limits $T\to0$ and $L\to\infty$ consequently do not commute.

The origin of this vanishing scale is especially transparent when computing the exact finite-size spectrum when $T\to 0$. The ground-state splitting is $\Delta E_L\equiv E_{0,\mathrm{PBC}}-E_{0,\mathrm{APBC}}=2t\tan[\pi/(2L)]=\pi v_{\rm F}/(2L)+O(L^{-3})$, with Fermi velocity $v_{\rm F}=2t$ at half filling; see the SM~\cite{SM}. Including the twofold degeneracy of the PBC ground state gives $\langle{\cal S}\rangle_{\rm B}\simeq2\exp(-\beta\Delta E_L)$ at sufficiently low temperature. Any crossover defined by a fixed value of the sign therefore obeys $T_\times\propto L^{-1}$, consistent both with the inset of Fig.~\ref{fig:exact-sign-1d}(a) and with the collapse as a function of $LT/v_{\rm F}$ in Fig.~\ref{fig:exact-sign-1d}(c).

This one-dimensional result establishes a finite-size crossover controlled by the gapless ground state, not a thermal phase transition. Indeed, finite-range one-dimensional quantum lattices have no transition at $T>0$~\cite{Araki1969}, whereas the half-filled hard-core-boson chain is a critical Luttinger liquid at $T=0$~\cite{Haldane1981}. This result thus provides a clean baseline for a two-dimensional case, where the hard-core-boson reference system instead undergoes a finite-temperature Berezinski-Kosterlitz-Thouless (BKT) transition.

\prlsection{Two dimensions} We now consider half-filled spinless fermions on an $N_s =L^2$ square lattice. In the world-line representation, sampling the absolute values of the fermionic weights generates the hard-core-boson reference ensemble with the same hopping, much as in the one-dimensional case. As before, the average sign is therefore the ratio of the fermionic and bosonic partition functions. While we have seen that in one dimension, the latter reduces to a free-fermion problem via the Jordan-Wigner transformation, unfortunately, no analogous local free-particle representation exists in two dimensions, and the hard-core-boson partition function must instead be sampled numerically. This requires no separate simulation: sign-free bosonic observables can be accumulated in the same run by omitting the permutation sign from the estimator.

\begin{figure}[t]
    \centering
    \includegraphics[width=\columnwidth]{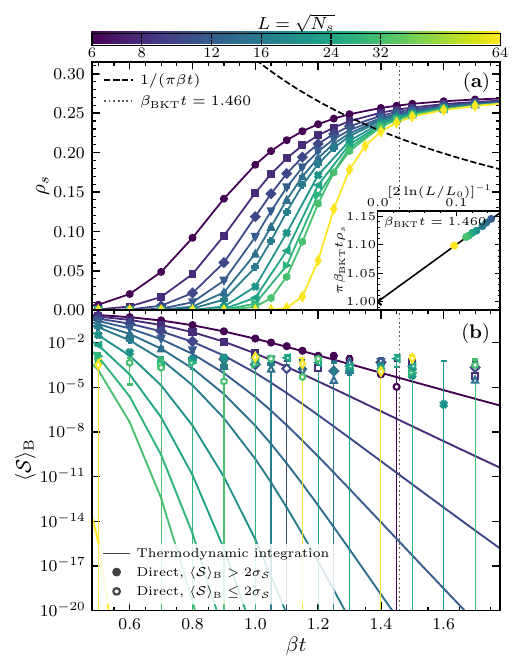}
    \caption{Half-filled square lattice at $t\Delta\tau=0.05$. (a) Hard-core-boson superfluid density from winding-number fluctuations. The dashed curve is the universal-jump condition $\rho_s=1/(\pi\beta t)$, and the inset shows the analysis of Eq.~\eqref{eq:weber-minnhagen} yielding $\beta_{\rm BKT}t\simeq1.460$. (b) Average fermion sign. Solid curves are reconstructed by thermodynamic integration; filled symbols are direct estimates resolved by more than two standard errors, whereas open symbols are unresolved. The dotted line marks the BKT estimate from panel (a).}
    \label{fig:sign-thermodynamic-integration}
\end{figure}

We now consider the consequences of this connection. Figure \ref{fig:sign-thermodynamic-integration}(a) shows the corresponding superfluid density, $\rho_s=\langle W_x^2+W_y^2\rangle_{\rm B}/(4\beta t)$, obtained from winding-number fluctuations~\cite{PollockCeperley1987}. Its intersection with the universal-jump condition $\rho_s=1/(\pi\beta t)$ identifies the vicinity of the BKT transition~\cite{NelsonKosterlitz1977}. Accounting for the leading logarithmic finite-size correction through the form~\cite{WeberMinnhagen1988,HaradaKawashima1997},
\begin{equation}
    \pi\beta_{\rm BKT}t\,\rho_s(\beta_{\rm BKT},L)
    =1+\frac{1}{2\ln (L/L_0)},
    \label{eq:weber-minnhagen}
\end{equation}
where $L_0$ is a nonuniversal scale entering the logarithmic finite-size correction, which gives $\beta_{\rm BKT}t\simeq1.460$ [inset of Fig.~\ref{fig:sign-thermodynamic-integration}(a)]. This value is in good agreement with the literature~\cite{Ding1992,HaradaKawashima1997}.

Turning back to the fermionic case, Fig.~\ref{fig:sign-thermodynamic-integration}(b) exposes the scale of the problem, with a statistical limitation of the direct sign estimator. Although $\langle{\cal S}\rangle_{\rm B}$ is measured at essentially no additional cost, it is the small difference between positive- and negative-sign probabilities. Its absolute uncertainty therefore remains finite as its mean decreases, and direct sampling ceases to resolve the sign once $\langle{\cal S}\rangle_{\rm B}\lesssim2\sigma_{\cal S}$, where $\sigma_{\cal S}$ denotes the standard error of the mean, effectively preventing one from using the average sign to quantify critical behavior.

Nonetheless, this is not an impediment if one recalls that we can reconstruct the same quantity directly from a free-energy difference. That is, given that $\langle {\cal S}\rangle_{\rm B}=\frac{{\cal Z}_{\rm F}}{{\cal Z}_{\rm B}}=\exp[-\beta N_s\Delta f]$, with $\Delta f=f_{\rm F}-f_{\rm B}$, where $f_{\rm F,B}$ are free-energy densities, rather than attempting to measure this exponentially small ratio directly, we introduce the {\it intensive} quantity
\begin{equation}
    g(\beta)\equiv\beta\Delta f
    =-\frac{1}{N_s}\ln\langle {\cal S}\rangle_{\rm B}=-\frac{1}{N_s}\left(\ln{\cal Z}_{\rm F}-\ln{\cal Z}_{\rm B}\right)\ .
    \label{eq:g-definition}
\end{equation}
Using the energy densities $e_\alpha=-N_s^{-1}\partial_\beta\ln{\cal Z}_\alpha$ for $\alpha={\rm F,B}$ gives
\begin{equation}
    \frac{dg}{d\beta}=e_{\rm F}-e_{\rm B}\ .
    \label{eq:thermodynamic-integration-derivative}
\end{equation}
At infinite temperature, the two canonical Hilbert spaces contain the same number of fixed-particle configurations, so ${\cal Z}_{\rm F}(0)={\cal Z}_{\rm B}(0)$ and $g(0)=0$. Thus, integrating Eq.~\eqref{eq:thermodynamic-integration-derivative}, it follows that
\begin{align}
    \beta\Delta f(\beta)
    =\int_0^\beta d\beta'\,
      [e_{\rm F}(\beta')-e_{\rm B}(\beta')],
    \label{eq:thermodynamic-integration}
\end{align}
and thus,
\begin{equation}
\langle {\cal S}\rangle_{\rm B}^{\rm TI}
    =\exp\!\left\{-N_s\int_0^\beta d\beta'\,
      [e_{\rm F}(\beta')-e_{\rm B}(\beta')]\right\}\ ,    \label{eq:integrated-sign}
\end{equation}
where the superscript `TI' emphasizes that the average sign is inferred from thermodynamic integration rather than measured directly. For the noninteracting fermions, $e_{\rm F}$ is obtained without stochastic sampling from a fixed-particle-number recursion over the single-particle spectrum, while $e_{\rm B}$ is measured in the sign-free WLQMC ensemble. Propagating the uncertainty of $e_{\rm B}$ through the integration gives the solid curves in Fig.~\ref{fig:sign-thermodynamic-integration}(b). Where direct estimates remain resolved, they provide a check of the reconstruction; at larger $L$ and lower temperatures, thermodynamic integration continues far below the direct sampling floor~\footnote{Note that the curves combine the exact, and thus continuum-time fermion energy with bosonic data at $t\Delta\tau=0.05$ and hence retain an $O(\Delta\tau^2)$ mismatch not included in the statistical bands. A more detailed comparison would require either the fermionic energy on the same checkerboard discretization or an extrapolation of the bosonic energies in $\Delta\tau^2$.}.

Notably, this construction does not remove the sign problem. It essentially replaces an exponentially ill-conditioned cancellation by an integration of intensive observables, while the factor $N_s$ in Eq.~\eqref{eq:integrated-sign} still amplifies any error in the integral. It does, however, give access to the free-energy information that is hidden once the direct estimator reaches its sampling floor.

With that in hand, we can now infer how to extract critical behavior from the average sign. For this, we note that for a two-dimensional thermal BKT transition, one can separate the regular and singular contributions to the free-energy density as~\cite{FisherBarber1972,Kosterlitz1974}
\begin{equation}
        f_{\rm B}(\tau,L)=f_{{\rm B},{\rm reg}}(\tau)
          +L^{-2}{\Phi}(u_L,y_L)+\ldots\ ,
    \label{eq:bkt-free-energy-scaling}
\end{equation}
with scaling variables, $u_L=\tau[\ln(L/L_0)]^2$, and $y_L=[\ln(L/L_0)]^{-1}$, for the reduced temperature $\tau=(T-T_{\rm BKT})/T_{\rm BKT}$. The variable $u_L$ follows from the essential divergence of the BKT correlation length, whereas $y_L$ accounts for the marginal logarithmic corrections. 
The form of $u_L$ is the analog of the scaling variable $\tau L^{1/\nu}$ at a second-order transition. Since the free-fermion free energy is analytic, one can use the relation of the free-energy density with the average sign to write,
\begin{equation}
    \ln\langle{\cal S}\rangle_{{\rm B},L}
     =\beta L^2 [f_{{\rm B},{\rm reg}}(\tau)-f_{\rm F}(\tau)]+\beta{\Phi}(u_L,y_L)+\ldots\ .
    \label{eq:bkt-sign-scaling}
\end{equation}
This shows that while the overall regular contribution is $O(L^2)$, the singular part is only $O(1)$.  One can eliminate the leading regular contribution by comparing system sizes $L$ and $2L$, at the same temperature, by defining the sign residual
\begin{align}
    R_L^{({\cal S})}(\beta)
    &\equiv \ln\langle{\cal S}\rangle_{{\rm B},L}
    -\frac{1}{4}\ln\langle{\cal S}\rangle_{{\rm B},2L}
    \nonumber\\
    &=\beta\left[{\Phi}(u_L,y_L)
    -\frac{1}{4}{\Phi}(u_{2L},y_{2L})\right]+\ldots\ .
    \label{eq:bkt-sign-residual}
\end{align}

\begin{figure}[t]
\centering
\includegraphics[width=0.99\columnwidth]{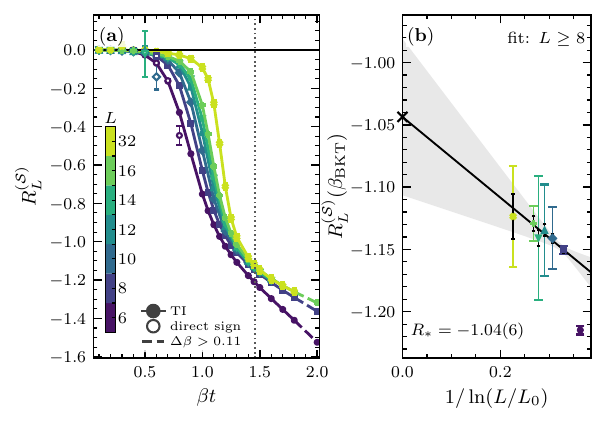}
\caption{Finite-size sign residuals $R_L^{({\cal S})}=\ln\langle{\cal S}\rangle_{{\rm B},L}-\tfrac14\ln\langle{\cal S}\rangle_{{\rm B},2L}$ obtained from thermodynamic integration at $t\Delta\tau=0.05$ where $\Delta\tau$ is the discretization interval of imaginary time.  See SM. (a) Inverse-temperature dependence for different size pairs $(L,2L)$. Filled symbols show the thermodynamic-integration estimates, while open symbols show direct-sign estimates where statistically resolved. The vertical dotted line marks the independent stiffness estimate $\beta_{\rm BKT}t=1.460$. (b) Residuals interpolated to $\beta_{\rm BKT}$ and plotted against $1/\ln(L/L_0)$. The solid line is a leading-logarithmic fit for $L\geq8$, with $L_0=0.375(14)$ fixed by the stiffness analysis [Fig.~\ref{fig:sign-thermodynamic-integration}], yielding $R_\ast=-1.04(7)$. The shaded band represents the one-standard-error fit uncertainty; colored and black error bars denote combined statistical-plus-quadrature and statistical-only uncertainties, respectively.}
\label{fig:kt_residual}
\end{figure}

Expanding the scaling function $\Phi$ in the running marginal field~\cite{Pelissetto2013}, one obtains at $T=T_{\rm BKT}$, to leading logarithmic order,
\begin{equation}
R_L^{(\mathcal S)}(\beta_{\mathrm{BKT}}) =
R_\ast+\frac{a_1}{\ln(L/L_0)}+\mathcal O\left[
\frac{\ln\ln(L/L_0)}
{\ln^2(L/L_0)}
\right],
\label{eq:residual}
\end{equation}
where $R_\ast=\frac34\beta_{\rm BKT}\Phi(0,0)$. 
Thus the residual has a finite critical limit that is approached only logarithmically. Indeed, Fig.~\ref{fig:kt_residual}(a) shows that the residuals become weakly size-dependent near the independently determined $\beta_{\rm BKT}$, whereas in  Fig.~\ref{fig:kt_residual}(b), their values at $\beta_{\rm BKT}$ for $L\geq8$ are consistent with the leading-logarithmic form in Eq.~\eqref{eq:residual}, yielding $R_\ast=-1.04(7)$.

\prlsection{Discussion and Outlook}
The examples investigated here extend earlier observations~\cite{Mondaini2022,Mou2022,Mondaini2023,SousaJunior2024ExtendedHubbard,Wang2024SignPAM,Yi2024} of the imprinting of phase transitions into the sign problem in two fundamental ways.  First, we consider the sign in the context of world-line QMC, as opposed to determinant QMC. Second, our study `inverts' previous analysis by considering phase transitions in the bosonic reference model when the (free) fermion model is completely analytic. It unequivocally shows how the average sign can be interpreted as a tracker of phase transitions, either in the original or reference models, so long as the non-analytic behaviors do not coincide in the space of parameters. An important point, though, is that in none of these cases does the scaling analysis rely directly on the data that comes from the statistical convergence of $\langle {\cal S}\rangle_{\rm B}$, but rather from the fact that the free energy of the fermionic system can be analytically extracted and the average sign inferred indirectly, supplanting the barrier that prevents one from obtaining the accurate averages of a quantity whose mean exponentially approaches zero.

Despite this limitation, the finite-size scaling framework developed here, particularly the residual combination that cancels the leading regular contribution to the free energy, can be adapted to other critical regimes, including those associated with {\it quantum} phase transitions. Candidates for this analysis in two dimensions are free fermions in the presence of a staggered potential $\hat {\cal H}_{\pm \Delta} \equiv \hat {\cal H}_{\rm F} + \Delta \sum_i (-1)^{x_i+y_i}\hat c_i^\dagger \hat c_i^{\phantom {\dagger}}$,  for which the reference model is hardcore bosons in the presence of the same lattice potential. Although the noninteracting fermions are already band insulating for any $\Delta\neq0$, their free energy remains analytic near the finite critical $\Delta$ of the reference system. The corresponding hard-core bosons instead undergo a continuous superfluid-to-insulator transition in the three-dimensional XY universality class~\cite{Priyadarshee2006,HenRigol2009,HenIskinRigol2010}. The average-sign residual should therefore inherit the bosonic critical contribution.

Another case is that of free fermions in the presence of disorder, $\hat{\mathcal H}_{W}\equiv\hat{\mathcal H}_{\rm F}+W\sum_i\epsilon_i\hat c_i^\dagger\hat c_i^{\phantom{\dagger}}$, where the dimensionless $\epsilon_i$ are independent random variables uniformly distributed on $[-1,1]$. The on-site potentials are therefore uniformly distributed on $[-W,W]$. In this case, noninteracting fermions in two dimensions are Anderson localized for arbitrarily weak $W$ in the thermodynamic limit, without a finite-$W$ thermodynamic transition~\cite{Abrahams1979}. The hard-core-boson reference system, on the other hand, remains superfluid at weak disorder and undergoes a transition at finite disorder to a gapless and compressible Bose-glass phase~\cite{Fisher1989,AlvarezZuniga2015,NgSorensen2015}. Near this transition, the fermionic free energy is expected to remain analytic while the bosonic contribution becomes singular, making the average sign a potential probe of the superfluid-Bose-glass critical point.

Both cases can be tackled numerically with QMC methods that efficiently connect winding-number sectors, such as continuous-time worm or operator-loop methods, which provide natural routes for future investigation~\cite{Prokofev1998,Sandvik1999}.

Our work highlights how general this approach is among any QMC family one chooses. That is, the same scaling analysis based on the cancellation of the regular part of the free energy carries over to other QMC formulations in which $\langle{\cal S}\rangle_{|W|}$ remains statistically accessible over a broader range of parameters. A natural setting is determinantal quantum Monte Carlo (DQMC)~\cite{Blankenbecler1981}, which is sign-problem-free in the noninteracting limit and for which finite-size scaling analyses of the average and spin-resolved signs have already revealed critical crossings and data collapse near interacting quantum critical points~\cite{Mondaini2022,Mondaini2023}. Extending the residual construction introduced here, with the size prefactor adjusted to the relevant space-time scaling, may help isolate the singular free-energy contribution even when the raw average sign is dominated by an extensive regular background. The physical interpretation will be less direct than in WLQMC, however, because the absolute-weight DQMC ensemble does not generally correspond to a simple independent quantum Hamiltonian, such as a direct bosonic version of the model. 

\prlsection{Acknowledgments} We acknowledge W. S. Oliveira and N. C. Costa for initial discussions around this project. We performed the numerical calculations on the Carya and Sabine clusters at the University of Houston's Research Computing Data Core. This work also used ACES at Texas A\&M High Performance Research Computing through allocation PHY240046 from the Advanced Cyberinfrastructure Coordination Ecosystem: Services \& Support (ACCESS) program, which is supported by U.S. National Science Foundation grants 2138259, 2138286, 2138307, 2137603, and 2138296. R.M.~acknowledges that the research was funded in part by the Robert A.~Welch Foundation, Grant \#L-E-0001-19921203. RTS is supported by the grant DOE DE-SC0014671 funded by the U.S. Department of Energy, Office of Science. The data that support the findings of this article are openly available~\cite{zenodo}.

\bibliography{references}

\begin{thebibliography}{47}%
\makeatletter
\providecommand \@ifxundefined [1]{%
 \@ifx{#1\undefined}
}%
\providecommand \@ifnum [1]{%
 \ifnum #1\expandafter \@firstoftwo
 \else \expandafter \@secondoftwo
 \fi
}%
\providecommand \@ifx [1]{%
 \ifx #1\expandafter \@firstoftwo
 \else \expandafter \@secondoftwo
 \fi
}%
\providecommand \natexlab [1]{#1}%
\providecommand \enquote  [1]{``#1''}%
\providecommand \bibnamefont  [1]{#1}%
\providecommand \bibfnamefont [1]{#1}%
\providecommand \citenamefont [1]{#1}%
\providecommand \href@noop [0]{\@secondoftwo}%
\providecommand \href [0]{\begingroup \@sanitize@url \@href}%
\providecommand \@href[1]{\@@startlink{#1}\@@href}%
\providecommand \@@href[1]{\endgroup#1\@@endlink}%
\providecommand \@sanitize@url [0]{\catcode `\\12\catcode `\$12\catcode
  `\&12\catcode `\#12\catcode `\^12\catcode `\_12\catcode `\%12\relax}%
\providecommand \@@startlink[1]{}%
\providecommand \@@endlink[0]{}%
\providecommand \url  [0]{\begingroup\@sanitize@url \@url }%
\providecommand \@url [1]{\endgroup\@href {#1}{\urlprefix }}%
\providecommand \urlprefix  [0]{URL }%
\providecommand \Eprint [0]{\href }%
\providecommand \doibase [0]{https://doi.org/}%
\providecommand \selectlanguage [0]{\@gobble}%
\providecommand \bibinfo  [0]{\@secondoftwo}%
\providecommand \bibfield  [0]{\@secondoftwo}%
\providecommand \translation [1]{[#1]}%
\providecommand \BibitemOpen [0]{}%
\providecommand \bibitemStop [0]{}%
\providecommand \bibitemNoStop [0]{.\EOS\space}%
\providecommand \EOS [0]{\spacefactor3000\relax}%
\providecommand \BibitemShut  [1]{\csname bibitem#1\endcsname}%
\let\auto@bib@innerbib\@empty
\bibitem [{\citenamefont {Foulkes}\ \emph {et~al.}(2001)\citenamefont
  {Foulkes}, \citenamefont {Mitas}, \citenamefont {Needs},\ and\ \citenamefont
  {Rajagopal}}]{Foulkes2001}%
  \BibitemOpen
  \bibfield  {author} {\bibinfo {author} {\bibfnamefont {W.~M.~C.}\
  \bibnamefont {Foulkes}}, \bibinfo {author} {\bibfnamefont {L.}~\bibnamefont
  {Mitas}}, \bibinfo {author} {\bibfnamefont {R.~J.}\ \bibnamefont {Needs}},\
  and\ \bibinfo {author} {\bibfnamefont {G.}~\bibnamefont {Rajagopal}},\
  }\bibfield  {title} {\bibinfo {title} {Quantum {Monte Carlo} simulations of
  solids},\ }\href {https://doi.org/10.1103/RevModPhys.73.33} {\bibfield
  {journal} {\bibinfo  {journal} {Rev. Mod. Phys.}\ }\textbf {\bibinfo {volume}
  {73}},\ \bibinfo {pages} {33} (\bibinfo {year} {2001})}\BibitemShut {NoStop}%
\bibitem [{\citenamefont {Austin}\ \emph {et~al.}(2012)\citenamefont {Austin},
  \citenamefont {Zubarev},\ and\ \citenamefont {Lester}}]{Austin2012QMC}%
  \BibitemOpen
  \bibfield  {author} {\bibinfo {author} {\bibfnamefont {B.~M.}\ \bibnamefont
  {Austin}}, \bibinfo {author} {\bibfnamefont {D.~Y.}\ \bibnamefont
  {Zubarev}},\ and\ \bibinfo {author} {\bibfnamefont {J.}~\bibnamefont
  {Lester}, \bibfnamefont {William~A.}},\ }\bibfield  {title} {\bibinfo {title}
  {Quantum {M}onte {C}arlo and related approaches},\ }\href
  {https://doi.org/10.1021/cr2001564} {\bibfield  {journal} {\bibinfo
  {journal} {Chemical Reviews}\ }\textbf {\bibinfo {volume} {112}},\ \bibinfo
  {pages} {263} (\bibinfo {year} {2012})}\BibitemShut {NoStop}%
\bibitem [{\citenamefont {Kaul}\ \emph {et~al.}(2013)\citenamefont {Kaul},
  \citenamefont {Melko},\ and\ \citenamefont {Sandvik}}]{Kaul2013QMC}%
  \BibitemOpen
  \bibfield  {author} {\bibinfo {author} {\bibfnamefont {R.~K.}\ \bibnamefont
  {Kaul}}, \bibinfo {author} {\bibfnamefont {R.~G.}\ \bibnamefont {Melko}},\
  and\ \bibinfo {author} {\bibfnamefont {A.~W.}\ \bibnamefont {Sandvik}},\
  }\bibfield  {title} {\bibinfo {title} {Bridging lattice-scale physics and
  continuum field theory with quantum {M}onte {C}arlo simulations},\ }\href
  {https://doi.org/10.1146/annurev-conmatphys-030212-184215} {\bibfield
  {journal} {\bibinfo  {journal} {Annual Review of Condensed Matter Physics}\
  }\textbf {\bibinfo {volume} {4}},\ \bibinfo {pages} {179} (\bibinfo {year}
  {2013})}\BibitemShut {NoStop}%
\bibitem [{\citenamefont {Carlson}\ \emph {et~al.}(2015)\citenamefont
  {Carlson}, \citenamefont {Gandolfi}, \citenamefont {Pederiva}, \citenamefont
  {Pieper}, \citenamefont {Schiavilla}, \citenamefont {Schmidt},\ and\
  \citenamefont {Wiringa}}]{Carlson2015NuclearQMC}%
  \BibitemOpen
  \bibfield  {author} {\bibinfo {author} {\bibfnamefont {J.}~\bibnamefont
  {Carlson}}, \bibinfo {author} {\bibfnamefont {S.}~\bibnamefont {Gandolfi}},
  \bibinfo {author} {\bibfnamefont {F.}~\bibnamefont {Pederiva}}, \bibinfo
  {author} {\bibfnamefont {S.~C.}\ \bibnamefont {Pieper}}, \bibinfo {author}
  {\bibfnamefont {R.}~\bibnamefont {Schiavilla}}, \bibinfo {author}
  {\bibfnamefont {K.~E.}\ \bibnamefont {Schmidt}},\ and\ \bibinfo {author}
  {\bibfnamefont {R.~B.}\ \bibnamefont {Wiringa}},\ }\bibfield  {title}
  {\bibinfo {title} {Quantum {M}onte {C}arlo methods for nuclear physics},\
  }\href {https://doi.org/10.1103/RevModPhys.87.1067} {\bibfield  {journal}
  {\bibinfo  {journal} {Reviews of Modern Physics}\ }\textbf {\bibinfo {volume}
  {87}},\ \bibinfo {pages} {1067} (\bibinfo {year} {2015})}\BibitemShut
  {NoStop}%
\bibitem [{\citenamefont {Li}\ and\ \citenamefont
  {Yao}(2019)}]{Li2019SignProblemFree}%
  \BibitemOpen
  \bibfield  {author} {\bibinfo {author} {\bibfnamefont {Z.-X.}\ \bibnamefont
  {Li}}\ and\ \bibinfo {author} {\bibfnamefont {H.}~\bibnamefont {Yao}},\
  }\bibfield  {title} {\bibinfo {title} {Sign-problem-free fermionic quantum
  {M}onte {C}arlo: Developments and applications},\ }\href
  {https://doi.org/10.1146/annurev-conmatphys-033117-054307} {\bibfield
  {journal} {\bibinfo  {journal} {Annual Review of Condensed Matter Physics}\
  }\textbf {\bibinfo {volume} {10}},\ \bibinfo {pages} {337} (\bibinfo {year}
  {2019})}\BibitemShut {NoStop}%
\bibitem [{\citenamefont {Nagata}(2022)}]{Nagata2022}%
  \BibitemOpen
  \bibfield  {author} {\bibinfo {author} {\bibfnamefont {K.}~\bibnamefont
  {Nagata}},\ }\bibfield  {title} {\bibinfo {title} {Finite-density lattice
  {QCD} and sign problem: {C}urrent status and open problems},\ }\href
  {https://doi.org/https://doi.org/10.1016/j.ppnp.2022.103991} {\bibfield
  {journal} {\bibinfo  {journal} {Progress in Particle and Nuclear Physics}\
  }\textbf {\bibinfo {volume} {127}},\ \bibinfo {pages} {103991} (\bibinfo
  {year} {2022})}\BibitemShut {NoStop}%
\bibitem [{\citenamefont {Gull}\ \emph {et~al.}(2011)\citenamefont {Gull},
  \citenamefont {Millis}, \citenamefont {Lichtenstein}, \citenamefont
  {Rubtsov}, \citenamefont {Troyer},\ and\ \citenamefont {Werner}}]{Gull2011}%
  \BibitemOpen
  \bibfield  {author} {\bibinfo {author} {\bibfnamefont {E.}~\bibnamefont
  {Gull}}, \bibinfo {author} {\bibfnamefont {A.~J.}\ \bibnamefont {Millis}},
  \bibinfo {author} {\bibfnamefont {A.~I.}\ \bibnamefont {Lichtenstein}},
  \bibinfo {author} {\bibfnamefont {A.~N.}\ \bibnamefont {Rubtsov}}, \bibinfo
  {author} {\bibfnamefont {M.}~\bibnamefont {Troyer}},\ and\ \bibinfo {author}
  {\bibfnamefont {P.}~\bibnamefont {Werner}},\ }\bibfield  {title} {\bibinfo
  {title} {Continuous-time {Monte Carlo} methods for quantum impurity models},\
  }\href {https://doi.org/10.1103/RevModPhys.83.349} {\bibfield  {journal}
  {\bibinfo  {journal} {Rev. Mod. Phys.}\ }\textbf {\bibinfo {volume} {83}},\
  \bibinfo {pages} {349} (\bibinfo {year} {2011})}\BibitemShut {NoStop}%
\bibitem [{\citenamefont {Loh}\ \emph {et~al.}(1990)\citenamefont {Loh},
  \citenamefont {Gubernatis}, \citenamefont {Scalettar}, \citenamefont {White},
  \citenamefont {Scalapino},\ and\ \citenamefont {Sugar}}]{Loh1990}%
  \BibitemOpen
  \bibfield  {author} {\bibinfo {author} {\bibfnamefont {E.~Y.}\ \bibnamefont
  {Loh}}, \bibinfo {author} {\bibfnamefont {J.~E.}\ \bibnamefont {Gubernatis}},
  \bibinfo {author} {\bibfnamefont {R.~T.}\ \bibnamefont {Scalettar}}, \bibinfo
  {author} {\bibfnamefont {S.~R.}\ \bibnamefont {White}}, \bibinfo {author}
  {\bibfnamefont {D.~J.}\ \bibnamefont {Scalapino}},\ and\ \bibinfo {author}
  {\bibfnamefont {R.~L.}\ \bibnamefont {Sugar}},\ }\bibfield  {title} {\bibinfo
  {title} {Sign problem in the numerical simulation of many-electron systems},\
  }\href {https://doi.org/10.1103/PhysRevB.41.9301} {\bibfield  {journal}
  {\bibinfo  {journal} {Phys. Rev. B}\ }\textbf {\bibinfo {volume} {41}},\
  \bibinfo {pages} {9301} (\bibinfo {year} {1990})}\BibitemShut {NoStop}%
\bibitem [{\citenamefont {Troyer}\ and\ \citenamefont
  {Wiese}(2005)}]{Troyer2005}%
  \BibitemOpen
  \bibfield  {author} {\bibinfo {author} {\bibfnamefont {M.}~\bibnamefont
  {Troyer}}\ and\ \bibinfo {author} {\bibfnamefont {U.-J.}\ \bibnamefont
  {Wiese}},\ }\bibfield  {title} {\bibinfo {title} {Computational complexity
  and fundamental limitations to fermionic quantum {Monte Carlo} simulations},\
  }\href {https://doi.org/10.1103/PhysRevLett.94.170201} {\bibfield  {journal}
  {\bibinfo  {journal} {Phys. Rev. Lett.}\ }\textbf {\bibinfo {volume} {94}},\
  \bibinfo {pages} {170201} (\bibinfo {year} {2005})}\BibitemShut {NoStop}%
\bibitem [{\citenamefont {Mondaini}\ \emph {et~al.}(2022)\citenamefont
  {Mondaini}, \citenamefont {Tarat},\ and\ \citenamefont
  {Scalettar}}]{Mondaini2022}%
  \BibitemOpen
  \bibfield  {author} {\bibinfo {author} {\bibfnamefont {R.}~\bibnamefont
  {Mondaini}}, \bibinfo {author} {\bibfnamefont {S.}~\bibnamefont {Tarat}},\
  and\ \bibinfo {author} {\bibfnamefont {R.~T.}\ \bibnamefont {Scalettar}},\
  }\bibfield  {title} {\bibinfo {title} {{Quantum critical points and the sign
  problem}},\ }\href {https://doi.org/10.1126/science.abg9299} {\bibfield
  {journal} {\bibinfo  {journal} {Science}\ }\textbf {\bibinfo {volume}
  {375}},\ \bibinfo {pages} {418} (\bibinfo {year} {2022})}\BibitemShut
  {NoStop}%
\bibitem [{\citenamefont {Mou}\ \emph {et~al.}(2022)\citenamefont {Mou},
  \citenamefont {Mondaini},\ and\ \citenamefont {Scalettar}}]{Mou2022}%
  \BibitemOpen
  \bibfield  {author} {\bibinfo {author} {\bibfnamefont {Y.}~\bibnamefont
  {Mou}}, \bibinfo {author} {\bibfnamefont {R.}~\bibnamefont {Mondaini}},\ and\
  \bibinfo {author} {\bibfnamefont {R.~T.}\ \bibnamefont {Scalettar}},\
  }\bibfield  {title} {\bibinfo {title} {Bilayer {H}ubbard model: Analysis
  based on the fermionic sign problem},\ }\href
  {https://doi.org/10.1103/PhysRevB.106.125116} {\bibfield  {journal} {\bibinfo
   {journal} {Phys. Rev. B}\ }\textbf {\bibinfo {volume} {106}},\ \bibinfo
  {pages} {125116} (\bibinfo {year} {2022})}\BibitemShut {NoStop}%
\bibitem [{\citenamefont {Mondaini}\ \emph {et~al.}(2023)\citenamefont
  {Mondaini}, \citenamefont {Tarat},\ and\ \citenamefont
  {Scalettar}}]{Mondaini2023}%
  \BibitemOpen
  \bibfield  {author} {\bibinfo {author} {\bibfnamefont {R.}~\bibnamefont
  {Mondaini}}, \bibinfo {author} {\bibfnamefont {S.}~\bibnamefont {Tarat}},\
  and\ \bibinfo {author} {\bibfnamefont {R.~T.}\ \bibnamefont {Scalettar}},\
  }\bibfield  {title} {\bibinfo {title} {{Universality and critical exponents
  of the fermion sign problem}},\ }\href
  {https://doi.org/10.1103/PhysRevB.107.245144} {\bibfield  {journal} {\bibinfo
   {journal} {Phys. Rev. B}\ }\textbf {\bibinfo {volume} {107}},\ \bibinfo
  {pages} {245144} (\bibinfo {year} {2023})}\BibitemShut {NoStop}%
\bibitem [{\citenamefont {Sousa-J{\'u}nior}\ \emph {et~al.}(2024)\citenamefont
  {Sousa-J{\'u}nior}, \citenamefont {Costa},\ and\ \citenamefont {dos
  Santos}}]{SousaJunior2024ExtendedHubbard}%
  \BibitemOpen
  \bibfield  {author} {\bibinfo {author} {\bibfnamefont {S.~d.~A.}\
  \bibnamefont {Sousa-J{\'u}nior}}, \bibinfo {author} {\bibfnamefont {N.~C.}\
  \bibnamefont {Costa}},\ and\ \bibinfo {author} {\bibfnamefont {R.~R.}\
  \bibnamefont {dos Santos}},\ }\bibfield  {title} {\bibinfo {title}
  {Half-filled extended {H}ubbard model on a square lattice: Phase boundaries
  from determinant quantum {M}onte {C}arlo simulations},\ }\href
  {https://doi.org/10.1103/PhysRevB.109.165102} {\bibfield  {journal} {\bibinfo
   {journal} {Physical Review B}\ }\textbf {\bibinfo {volume} {109}},\ \bibinfo
  {pages} {165102} (\bibinfo {year} {2024})}\BibitemShut {NoStop}%
\bibitem [{\citenamefont {Wang}\ \emph {et~al.}(2024)\citenamefont {Wang},
  \citenamefont {Hu},\ and\ \citenamefont {Yang}}]{Wang2024SignPAM}%
  \BibitemOpen
  \bibfield  {author} {\bibinfo {author} {\bibfnamefont {M.}~\bibnamefont
  {Wang}}, \bibinfo {author} {\bibfnamefont {D.}~\bibnamefont {Hu}},\ and\
  \bibinfo {author} {\bibfnamefont {Y.-f.}\ \bibnamefont {Yang}},\ }\bibfield
  {title} {\bibinfo {title} {Insulator-to-insulator transition and sign problem
  in the periodic {A}nderson model with a staggered potential},\ }\href
  {https://doi.org/10.1103/PhysRevB.110.195122} {\bibfield  {journal} {\bibinfo
   {journal} {Physical Review B}\ }\textbf {\bibinfo {volume} {110}},\ \bibinfo
  {pages} {195122} (\bibinfo {year} {2024})}\BibitemShut {NoStop}%
\bibitem [{\citenamefont {Yi}\ \emph {et~al.}(2024)\citenamefont {Yi},
  \citenamefont {Cheng}, \citenamefont {Pil\'e}, \citenamefont {Burovski},\
  and\ \citenamefont {Mondaini}}]{Yi2024}%
  \BibitemOpen
  \bibfield  {author} {\bibinfo {author} {\bibfnamefont {T.-C.}\ \bibnamefont
  {Yi}}, \bibinfo {author} {\bibfnamefont {S.}~\bibnamefont {Cheng}}, \bibinfo
  {author} {\bibfnamefont {I.}~\bibnamefont {Pil\'e}}, \bibinfo {author}
  {\bibfnamefont {E.}~\bibnamefont {Burovski}},\ and\ \bibinfo {author}
  {\bibfnamefont {R.}~\bibnamefont {Mondaini}},\ }\bibfield  {title} {\bibinfo
  {title} {Two-dimensional polarized superfluids through the prism of the
  fermion sign problem},\ }\href {https://doi.org/10.1103/PhysRevB.110.085131}
  {\bibfield  {journal} {\bibinfo  {journal} {Phys. Rev. B}\ }\textbf {\bibinfo
  {volume} {110}},\ \bibinfo {pages} {085131} (\bibinfo {year}
  {2024})}\BibitemShut {NoStop}%
\bibitem [{\citenamefont {Hirsch}\ \emph {et~al.}(1981)\citenamefont {Hirsch},
  \citenamefont {Scalapino}, \citenamefont {Sugar},\ and\ \citenamefont
  {Blankenbecler}}]{Hirsch1981}%
  \BibitemOpen
  \bibfield  {author} {\bibinfo {author} {\bibfnamefont {J.~E.}\ \bibnamefont
  {Hirsch}}, \bibinfo {author} {\bibfnamefont {D.~J.}\ \bibnamefont
  {Scalapino}}, \bibinfo {author} {\bibfnamefont {R.~L.}\ \bibnamefont
  {Sugar}},\ and\ \bibinfo {author} {\bibfnamefont {R.}~\bibnamefont
  {Blankenbecler}},\ }\bibfield  {title} {\bibinfo {title} {Efficient {M}onte
  {C}arlo procedure for systems with fermions},\ }\href
  {https://doi.org/10.1103/PhysRevLett.47.1628} {\bibfield  {journal} {\bibinfo
   {journal} {Phys. Rev. Lett.}\ }\textbf {\bibinfo {volume} {47}},\ \bibinfo
  {pages} {1628} (\bibinfo {year} {1981})}\BibitemShut {NoStop}%
\bibitem [{\citenamefont {Hirsch}\ \emph {et~al.}(1982)\citenamefont {Hirsch},
  \citenamefont {Sugar}, \citenamefont {Scalapino},\ and\ \citenamefont
  {Blankenbecler}}]{Hirsch1982}%
  \BibitemOpen
  \bibfield  {author} {\bibinfo {author} {\bibfnamefont {J.~E.}\ \bibnamefont
  {Hirsch}}, \bibinfo {author} {\bibfnamefont {R.~L.}\ \bibnamefont {Sugar}},
  \bibinfo {author} {\bibfnamefont {D.~J.}\ \bibnamefont {Scalapino}},\ and\
  \bibinfo {author} {\bibfnamefont {R.}~\bibnamefont {Blankenbecler}},\
  }\bibfield  {title} {\bibinfo {title} {Monte {C}arlo simulations of
  one-dimensional fermion systems},\ }\href
  {https://doi.org/10.1103/PhysRevB.26.5033} {\bibfield  {journal} {\bibinfo
  {journal} {Phys. Rev. B}\ }\textbf {\bibinfo {volume} {26}},\ \bibinfo
  {pages} {5033} (\bibinfo {year} {1982})}\BibitemShut {NoStop}%
\bibitem [{\citenamefont {Scalettar}(1999)}]{Scalettar1999}%
  \BibitemOpen
  \bibfield  {author} {\bibinfo {author} {\bibfnamefont {R.~T.}\ \bibnamefont
  {Scalettar}},\ }\bibfield  {title} {\bibinfo {title} {World-line quantum
  {M}onte {C}arlo},\ }in\ \href {https://doi.org/10.1007/978-94-011-4792-7_3}
  {\emph {\bibinfo {booktitle} {Quantum Monte Carlo Methods in Physics and
  Chemistry}}},\ \bibinfo {editor} {edited by\ \bibinfo {editor} {\bibfnamefont
  {M.~P.}\ \bibnamefont {Nightingale}}\ and\ \bibinfo {editor} {\bibfnamefont
  {C.~J.}\ \bibnamefont {Umrigar}}}\ (\bibinfo  {publisher} {Springer},\
  \bibinfo {address} {Dordrecht},\ \bibinfo {year} {1999})\ pp.\ \bibinfo
  {pages} {65--100}\BibitemShut {NoStop}%
\bibitem [{\citenamefont {Assaad}\ and\ \citenamefont
  {Evertz}(2008)}]{AssaadEvertz2008}%
  \BibitemOpen
  \bibfield  {author} {\bibinfo {author} {\bibfnamefont {F.~F.}\ \bibnamefont
  {Assaad}}\ and\ \bibinfo {author} {\bibfnamefont {H.~G.}\ \bibnamefont
  {Evertz}},\ }\bibfield  {title} {\bibinfo {title} {World-line and
  determinantal quantum {M}onte {C}arlo methods for spins, phonons and
  electrons},\ }in\ \href {https://doi.org/10.1007/978-3-540-74686-7_10} {\emph
  {\bibinfo {booktitle} {Computational Many-Particle Physics}}},\ \bibinfo
  {series} {Lecture Notes in Physics}, Vol.\ \bibinfo {volume} {739},\ \bibinfo
  {editor} {edited by\ \bibinfo {editor} {\bibfnamefont {H.}~\bibnamefont
  {Fehske}}, \bibinfo {editor} {\bibfnamefont {R.}~\bibnamefont {Schneider}},\
  and\ \bibinfo {editor} {\bibfnamefont {A.}~\bibnamefont {Wei{\ss}e}}}\
  (\bibinfo  {publisher} {Springer},\ \bibinfo {address} {Berlin},\ \bibinfo
  {year} {2008})\ pp.\ \bibinfo {pages} {277--356}\BibitemShut {NoStop}%
\bibitem [{Note1()}]{Note1}%
  \BibitemOpen
  \bibinfo {note} {In one dimension, there is a small but important algorithmic
  point. With local world-line updates, the spatial winding number cannot
  change. A simulation started from straight world lines therefore stays in the
  zero-winding sector, where the fermion sign is positive. Once loop updates
  are allowed, this is no longer true: a world line can wind around the ring,
  giving a cyclic permutation and, for even particle number, a negative
  contribution.}\BibitemShut {Stop}%
\bibitem [{\citenamefont {Jordan}\ and\ \citenamefont
  {Wigner}(1928)}]{Jordan1928}%
  \BibitemOpen
  \bibfield  {author} {\bibinfo {author} {\bibfnamefont {P.}~\bibnamefont
  {Jordan}}\ and\ \bibinfo {author} {\bibfnamefont {E.}~\bibnamefont
  {Wigner}},\ }\bibfield  {title} {\bibinfo {title} {{\"U}ber das paulische
  {\"a}quivalenzverbot},\ }\href {https://doi.org/10.1007/BF01331938}
  {\bibfield  {journal} {\bibinfo  {journal} {Zeitschrift f{\"u}r Physik}\
  }\textbf {\bibinfo {volume} {47}},\ \bibinfo {pages} {631} (\bibinfo {year}
  {1928})}\BibitemShut {NoStop}%
\bibitem [{\citenamefont {Lieb}\ \emph {et~al.}(1961)\citenamefont {Lieb},
  \citenamefont {Schultz},\ and\ \citenamefont {Mattis}}]{Lieb1961}%
  \BibitemOpen
  \bibfield  {author} {\bibinfo {author} {\bibfnamefont {E.}~\bibnamefont
  {Lieb}}, \bibinfo {author} {\bibfnamefont {T.}~\bibnamefont {Schultz}},\ and\
  \bibinfo {author} {\bibfnamefont {D.}~\bibnamefont {Mattis}},\ }\bibfield
  {title} {\bibinfo {title} {Two soluble models of an antiferromagnetic
  chain},\ }\href {https://doi.org/10.1016/0003-4916(61)90115-4} {\bibfield
  {journal} {\bibinfo  {journal} {Ann. Phys. (N.Y.)}\ }\textbf {\bibinfo
  {volume} {16}},\ \bibinfo {pages} {407} (\bibinfo {year} {1961})}\BibitemShut
  {NoStop}%
\bibitem [{SM()}]{SM}%
  \BibitemOpen
  \href@noop {} {\bibinfo {title} {Supplemental material for ``{Sign problem
  and criticality in world-line quantum Monte Carlo methods}''}},\ \bibinfo
  {note} {contains additional derivations, numerical details, and supporting
  results}\BibitemShut {NoStop}%
\bibitem [{\citenamefont {Araki}(1969)}]{Araki1969}%
  \BibitemOpen
  \bibfield  {author} {\bibinfo {author} {\bibfnamefont {H.}~\bibnamefont
  {Araki}},\ }\bibfield  {title} {\bibinfo {title} {Gibbs states of a one
  dimensional quantum lattice},\ }\href {https://doi.org/10.1007/BF01645134}
  {\bibfield  {journal} {\bibinfo  {journal} {Commun. Math. Phys.}\ }\textbf
  {\bibinfo {volume} {14}},\ \bibinfo {pages} {120} (\bibinfo {year}
  {1969})}\BibitemShut {NoStop}%
\bibitem [{\citenamefont {Haldane}(1981)}]{Haldane1981}%
  \BibitemOpen
  \bibfield  {author} {\bibinfo {author} {\bibfnamefont {F.~D.~M.}\
  \bibnamefont {Haldane}},\ }\bibfield  {title} {\bibinfo {title} {{Luttinger
  liquid theory} of one-dimensional quantum fluids. {I}. properties of the
  {Luttinger} model and their extension to the general {1D} interacting
  spinless {Fermi} gas},\ }\href {https://doi.org/10.1088/0022-3719/14/19/010}
  {\bibfield  {journal} {\bibinfo  {journal} {J. Phys. C: Solid State Phys.}\
  }\textbf {\bibinfo {volume} {14}},\ \bibinfo {pages} {2585} (\bibinfo {year}
  {1981})}\BibitemShut {NoStop}%
\bibitem [{\citenamefont {Pollock}\ and\ \citenamefont
  {Ceperley}(1987)}]{PollockCeperley1987}%
  \BibitemOpen
  \bibfield  {author} {\bibinfo {author} {\bibfnamefont {E.~L.}\ \bibnamefont
  {Pollock}}\ and\ \bibinfo {author} {\bibfnamefont {D.~M.}\ \bibnamefont
  {Ceperley}},\ }\bibfield  {title} {\bibinfo {title} {Path-integral
  computation of superfluid densities},\ }\href
  {https://doi.org/10.1103/PhysRevB.36.8343} {\bibfield  {journal} {\bibinfo
  {journal} {Phys. Rev. B}\ }\textbf {\bibinfo {volume} {36}},\ \bibinfo
  {pages} {8343} (\bibinfo {year} {1987})}\BibitemShut {NoStop}%
\bibitem [{\citenamefont {Nelson}\ and\ \citenamefont
  {Kosterlitz}(1977)}]{NelsonKosterlitz1977}%
  \BibitemOpen
  \bibfield  {author} {\bibinfo {author} {\bibfnamefont {D.~R.}\ \bibnamefont
  {Nelson}}\ and\ \bibinfo {author} {\bibfnamefont {J.~M.}\ \bibnamefont
  {Kosterlitz}},\ }\bibfield  {title} {\bibinfo {title} {Universal jump in the
  superfluid density of two-dimensional superfluids},\ }\href
  {https://doi.org/10.1103/PhysRevLett.39.1201} {\bibfield  {journal} {\bibinfo
   {journal} {Phys. Rev. Lett.}\ }\textbf {\bibinfo {volume} {39}},\ \bibinfo
  {pages} {1201} (\bibinfo {year} {1977})}\BibitemShut {NoStop}%
\bibitem [{\citenamefont {Weber}\ and\ \citenamefont
  {Minnhagen}(1988)}]{WeberMinnhagen1988}%
  \BibitemOpen
  \bibfield  {author} {\bibinfo {author} {\bibfnamefont {H.}~\bibnamefont
  {Weber}}\ and\ \bibinfo {author} {\bibfnamefont {P.}~\bibnamefont
  {Minnhagen}},\ }\bibfield  {title} {\bibinfo {title} {Monte {C}arlo
  determination of the critical temperature for the two-dimensional {$XY$}
  model},\ }\href {https://doi.org/10.1103/PhysRevB.37.5986} {\bibfield
  {journal} {\bibinfo  {journal} {Phys. Rev. B}\ }\textbf {\bibinfo {volume}
  {37}},\ \bibinfo {pages} {5986} (\bibinfo {year} {1988})}\BibitemShut
  {NoStop}%
\bibitem [{\citenamefont {Harada}\ and\ \citenamefont
  {Kawashima}(1997)}]{HaradaKawashima1997}%
  \BibitemOpen
  \bibfield  {author} {\bibinfo {author} {\bibfnamefont {K.}~\bibnamefont
  {Harada}}\ and\ \bibinfo {author} {\bibfnamefont {N.}~\bibnamefont
  {Kawashima}},\ }\bibfield  {title} {\bibinfo {title} {Universal jump in the
  helicity modulus of the two-dimensional quantum {$XY$} model},\ }\href
  {https://doi.org/10.1103/PhysRevB.55.R11949} {\bibfield  {journal} {\bibinfo
  {journal} {Phys. Rev. B}\ }\textbf {\bibinfo {volume} {55}},\ \bibinfo
  {pages} {R11949} (\bibinfo {year} {1997})}\BibitemShut {NoStop}%
\bibitem [{\citenamefont {Ding}(1992)}]{Ding1992}%
  \BibitemOpen
  \bibfield  {author} {\bibinfo {author} {\bibfnamefont {H.-Q.}\ \bibnamefont
  {Ding}},\ }\bibfield  {title} {\bibinfo {title} {Phase transition and
  thermodynamics of quantum xy model in two dimensions},\ }\href
  {https://doi.org/10.1103/PhysRevB.45.230} {\bibfield  {journal} {\bibinfo
  {journal} {Physical Review B}\ }\textbf {\bibinfo {volume} {45}},\ \bibinfo
  {pages} {230} (\bibinfo {year} {1992})}\BibitemShut {NoStop}%
\bibitem [{Note2()}]{Note2}%
  \BibitemOpen
  \bibinfo {note} {Note that the curves combine the exact, and thus
  continuum-time fermion energy with bosonic data at $t\Delta \tau =0.05$ and
  hence retain an $O(\Delta \tau ^2)$ mismatch not included in the statistical
  bands. A more detailed comparison would require either the fermionic energy
  on the same checkerboard discretization or an extrapolation of the bosonic
  energies in $\Delta \tau ^2$.}\BibitemShut {Stop}%
\bibitem [{\citenamefont {Fisher}\ and\ \citenamefont
  {Barber}(1972)}]{FisherBarber1972}%
  \BibitemOpen
  \bibfield  {author} {\bibinfo {author} {\bibfnamefont {M.~E.}\ \bibnamefont
  {Fisher}}\ and\ \bibinfo {author} {\bibfnamefont {M.~N.}\ \bibnamefont
  {Barber}},\ }\bibfield  {title} {\bibinfo {title} {Scaling theory for
  finite-size effects in the critical region},\ }\href
  {https://doi.org/10.1103/PhysRevLett.28.1516} {\bibfield  {journal} {\bibinfo
   {journal} {Phys. Rev. Lett.}\ }\textbf {\bibinfo {volume} {28}},\ \bibinfo
  {pages} {1516} (\bibinfo {year} {1972})}\BibitemShut {NoStop}%
\bibitem [{\citenamefont {Kosterlitz}(1974)}]{Kosterlitz1974}%
  \BibitemOpen
  \bibfield  {author} {\bibinfo {author} {\bibfnamefont {J.~M.}\ \bibnamefont
  {Kosterlitz}},\ }\bibfield  {title} {\bibinfo {title} {The critical
  properties of the two-dimensional {$XY$} model},\ }\href
  {https://doi.org/10.1088/0022-3719/7/6/005} {\bibfield  {journal} {\bibinfo
  {journal} {J. Phys. C: Solid State Phys.}\ }\textbf {\bibinfo {volume} {7}},\
  \bibinfo {pages} {1046} (\bibinfo {year} {1974})}\BibitemShut {NoStop}%
\bibitem [{\citenamefont {Pelissetto}\ and\ \citenamefont
  {Vicari}(2013)}]{Pelissetto2013}%
  \BibitemOpen
  \bibfield  {author} {\bibinfo {author} {\bibfnamefont {A.}~\bibnamefont
  {Pelissetto}}\ and\ \bibinfo {author} {\bibfnamefont {E.}~\bibnamefont
  {Vicari}},\ }\bibfield  {title} {\bibinfo {title} {Renormalization-group flow
  and asymptotic behaviors at the {Berezinskii-Kosterlitz-Thouless}
  transitions},\ }\href {https://doi.org/10.1103/PhysRevE.87.032105} {\bibfield
   {journal} {\bibinfo  {journal} {Phys. Rev. E}\ }\textbf {\bibinfo {volume}
  {87}},\ \bibinfo {pages} {032105} (\bibinfo {year} {2013})}\BibitemShut
  {NoStop}%
\bibitem [{\citenamefont {Priyadarshee}\ \emph {et~al.}(2006)\citenamefont
  {Priyadarshee}, \citenamefont {Chandrasekharan}, \citenamefont {Lee},\ and\
  \citenamefont {Baranger}}]{Priyadarshee2006}%
  \BibitemOpen
  \bibfield  {author} {\bibinfo {author} {\bibfnamefont {A.}~\bibnamefont
  {Priyadarshee}}, \bibinfo {author} {\bibfnamefont {S.}~\bibnamefont
  {Chandrasekharan}}, \bibinfo {author} {\bibfnamefont {J.-W.}\ \bibnamefont
  {Lee}},\ and\ \bibinfo {author} {\bibfnamefont {H.~U.}\ \bibnamefont
  {Baranger}},\ }\bibfield  {title} {\bibinfo {title} {Quantum phase
  transitions of hard-core bosons in background potentials},\ }\href
  {https://doi.org/10.1103/PhysRevLett.97.115703} {\bibfield  {journal}
  {\bibinfo  {journal} {Phys. Rev. Lett.}\ }\textbf {\bibinfo {volume} {97}},\
  \bibinfo {pages} {115703} (\bibinfo {year} {2006})}\BibitemShut {NoStop}%
\bibitem [{\citenamefont {Hen}\ and\ \citenamefont
  {Rigol}(2009)}]{HenRigol2009}%
  \BibitemOpen
  \bibfield  {author} {\bibinfo {author} {\bibfnamefont {I.}~\bibnamefont
  {Hen}}\ and\ \bibinfo {author} {\bibfnamefont {M.}~\bibnamefont {Rigol}},\
  }\bibfield  {title} {\bibinfo {title} {Superfluid to {M}ott-insulator
  transition of hardcore bosons in a superlattice},\ }\href
  {https://doi.org/10.1103/PhysRevB.80.134508} {\bibfield  {journal} {\bibinfo
  {journal} {Phys. Rev. B}\ }\textbf {\bibinfo {volume} {80}},\ \bibinfo
  {pages} {134508} (\bibinfo {year} {2009})}\BibitemShut {NoStop}%
\bibitem [{\citenamefont {Hen}\ \emph {et~al.}(2010)\citenamefont {Hen},
  \citenamefont {Iskin},\ and\ \citenamefont {Rigol}}]{HenIskinRigol2010}%
  \BibitemOpen
  \bibfield  {author} {\bibinfo {author} {\bibfnamefont {I.}~\bibnamefont
  {Hen}}, \bibinfo {author} {\bibfnamefont {M.}~\bibnamefont {Iskin}},\ and\
  \bibinfo {author} {\bibfnamefont {M.}~\bibnamefont {Rigol}},\ }\bibfield
  {title} {\bibinfo {title} {Phase diagram of the hard-core {B}ose-{H}ubbard
  model on a checkerboard superlattice},\ }\href
  {https://doi.org/10.1103/PhysRevB.81.064503} {\bibfield  {journal} {\bibinfo
  {journal} {Phys. Rev. B}\ }\textbf {\bibinfo {volume} {81}},\ \bibinfo
  {pages} {064503} (\bibinfo {year} {2010})}\BibitemShut {NoStop}%
\bibitem [{\citenamefont {Abrahams}\ \emph {et~al.}(1979)\citenamefont
  {Abrahams}, \citenamefont {Anderson}, \citenamefont {Licciardello},\ and\
  \citenamefont {Ramakrishnan}}]{Abrahams1979}%
  \BibitemOpen
  \bibfield  {author} {\bibinfo {author} {\bibfnamefont {E.}~\bibnamefont
  {Abrahams}}, \bibinfo {author} {\bibfnamefont {P.~W.}\ \bibnamefont
  {Anderson}}, \bibinfo {author} {\bibfnamefont {D.~C.}\ \bibnamefont
  {Licciardello}},\ and\ \bibinfo {author} {\bibfnamefont {T.~V.}\ \bibnamefont
  {Ramakrishnan}},\ }\bibfield  {title} {\bibinfo {title} {Scaling theory of
  localization: Absence of quantum diffusion in two dimensions},\ }\href
  {https://doi.org/10.1103/PhysRevLett.42.673} {\bibfield  {journal} {\bibinfo
  {journal} {Phys. Rev. Lett.}\ }\textbf {\bibinfo {volume} {42}},\ \bibinfo
  {pages} {673} (\bibinfo {year} {1979})}\BibitemShut {NoStop}%
\bibitem [{\citenamefont {Fisher}\ \emph {et~al.}(1989)\citenamefont {Fisher},
  \citenamefont {Weichman}, \citenamefont {Grinstein},\ and\ \citenamefont
  {Fisher}}]{Fisher1989}%
  \BibitemOpen
  \bibfield  {author} {\bibinfo {author} {\bibfnamefont {M.~P.~A.}\
  \bibnamefont {Fisher}}, \bibinfo {author} {\bibfnamefont {P.~B.}\
  \bibnamefont {Weichman}}, \bibinfo {author} {\bibfnamefont {G.}~\bibnamefont
  {Grinstein}},\ and\ \bibinfo {author} {\bibfnamefont {D.~S.}\ \bibnamefont
  {Fisher}},\ }\bibfield  {title} {\bibinfo {title} {Boson localization and the
  superfluid-insulator transition},\ }\href
  {https://doi.org/10.1103/PhysRevB.40.546} {\bibfield  {journal} {\bibinfo
  {journal} {Phys. Rev. B}\ }\textbf {\bibinfo {volume} {40}},\ \bibinfo
  {pages} {546} (\bibinfo {year} {1989})}\BibitemShut {NoStop}%
\bibitem [{\citenamefont {{\'A}lvarez~Z{\'u}{\~n}iga}\ \emph
  {et~al.}(2015)\citenamefont {{\'A}lvarez~Z{\'u}{\~n}iga}, \citenamefont
  {Luitz}, \citenamefont {Lemari{\'e}},\ and\ \citenamefont
  {Laflorencie}}]{AlvarezZuniga2015}%
  \BibitemOpen
  \bibfield  {author} {\bibinfo {author} {\bibfnamefont {J.~P.}\ \bibnamefont
  {{\'A}lvarez~Z{\'u}{\~n}iga}}, \bibinfo {author} {\bibfnamefont {D.~J.}\
  \bibnamefont {Luitz}}, \bibinfo {author} {\bibfnamefont {G.}~\bibnamefont
  {Lemari{\'e}}},\ and\ \bibinfo {author} {\bibfnamefont {N.}~\bibnamefont
  {Laflorencie}},\ }\bibfield  {title} {\bibinfo {title} {Critical properties
  of the superfluid--{B}ose-glass transition in two dimensions},\ }\href
  {https://doi.org/10.1103/PhysRevLett.114.155301} {\bibfield  {journal}
  {\bibinfo  {journal} {Phys. Rev. Lett.}\ }\textbf {\bibinfo {volume} {114}},\
  \bibinfo {pages} {155301} (\bibinfo {year} {2015})}\BibitemShut {NoStop}%
\bibitem [{\citenamefont {Ng}\ and\ \citenamefont
  {S{\o}rensen}(2015)}]{NgSorensen2015}%
  \BibitemOpen
  \bibfield  {author} {\bibinfo {author} {\bibfnamefont {R.}~\bibnamefont
  {Ng}}\ and\ \bibinfo {author} {\bibfnamefont {E.~S.}\ \bibnamefont
  {S{\o}rensen}},\ }\bibfield  {title} {\bibinfo {title} {Quantum critical
  scaling of dirty bosons in two dimensions},\ }\href
  {https://doi.org/10.1103/PhysRevLett.114.255701} {\bibfield  {journal}
  {\bibinfo  {journal} {Phys. Rev. Lett.}\ }\textbf {\bibinfo {volume} {114}},\
  \bibinfo {pages} {255701} (\bibinfo {year} {2015})}\BibitemShut {NoStop}%
\bibitem [{\citenamefont {Prokof'ev}\ \emph {et~al.}(1998)\citenamefont
  {Prokof'ev}, \citenamefont {Svistunov},\ and\ \citenamefont
  {Tupitsyn}}]{Prokofev1998}%
  \BibitemOpen
  \bibfield  {author} {\bibinfo {author} {\bibfnamefont {N.~V.}\ \bibnamefont
  {Prokof'ev}}, \bibinfo {author} {\bibfnamefont {B.~V.}\ \bibnamefont
  {Svistunov}},\ and\ \bibinfo {author} {\bibfnamefont {I.~S.}\ \bibnamefont
  {Tupitsyn}},\ }\bibfield  {title} {\bibinfo {title} {``worm'' algorithm in
  quantum {M}onte {C}arlo simulations},\ }\href
  {https://doi.org/10.1016/S0375-9601(97)00957-2} {\bibfield  {journal}
  {\bibinfo  {journal} {Phys. Lett. A}\ }\textbf {\bibinfo {volume} {238}},\
  \bibinfo {pages} {253} (\bibinfo {year} {1998})}\BibitemShut {NoStop}%
\bibitem [{\citenamefont {Sandvik}(1999)}]{Sandvik1999}%
  \BibitemOpen
  \bibfield  {author} {\bibinfo {author} {\bibfnamefont {A.~W.}\ \bibnamefont
  {Sandvik}},\ }\bibfield  {title} {\bibinfo {title} {Stochastic series
  expansion method with operator-loop update},\ }\href
  {https://doi.org/10.1103/PhysRevB.59.R14157} {\bibfield  {journal} {\bibinfo
  {journal} {Phys. Rev. B}\ }\textbf {\bibinfo {volume} {59}},\ \bibinfo
  {pages} {R14157} (\bibinfo {year} {1999})}\BibitemShut {NoStop}%
\bibitem [{\citenamefont {Blankenbecler}\ \emph {et~al.}(1981)\citenamefont
  {Blankenbecler}, \citenamefont {Scalapino},\ and\ \citenamefont
  {Sugar}}]{Blankenbecler1981}%
  \BibitemOpen
  \bibfield  {author} {\bibinfo {author} {\bibfnamefont {R.}~\bibnamefont
  {Blankenbecler}}, \bibinfo {author} {\bibfnamefont {D.~J.}\ \bibnamefont
  {Scalapino}},\ and\ \bibinfo {author} {\bibfnamefont {R.~L.}\ \bibnamefont
  {Sugar}},\ }\bibfield  {title} {\bibinfo {title} {Monte {C}arlo calculations
  of coupled boson-fermion systems. {I}},\ }\href
  {https://doi.org/10.1103/PhysRevD.24.2278} {\bibfield  {journal} {\bibinfo
  {journal} {Phys. Rev. D}\ }\textbf {\bibinfo {volume} {24}},\ \bibinfo
  {pages} {2278} (\bibinfo {year} {1981})}\BibitemShut {NoStop}%
\bibitem [{\citenamefont {Mondaini}\ and\ \citenamefont
  {Scalettar}(2026)}]{zenodo}%
  \BibitemOpen
  \bibfield  {author} {\bibinfo {author} {\bibfnamefont {R.}~\bibnamefont
  {Mondaini}}\ and\ \bibinfo {author} {\bibfnamefont {R.}~\bibnamefont
  {Scalettar}},\ }\bibfield  {title} {\bibinfo {title} {Dataset for ``{S}ign
  problem and criticality in world-line quantum {Monte Carlo} methods''},\
  }\href {https://doi.org/10.5281/zenodo.23004608} {10.5281/zenodo.23004608}
  (\bibinfo {year} {2026})\BibitemShut {NoStop}%
\bibitem [{\citenamefont {Rousseau}\ \emph {et~al.}(2005)\citenamefont
  {Rousseau}, \citenamefont {Scalettar},\ and\ \citenamefont
  {Batrouni}}]{Rousseau2005}%
  \BibitemOpen
  \bibfield  {author} {\bibinfo {author} {\bibfnamefont {V.~G.}\ \bibnamefont
  {Rousseau}}, \bibinfo {author} {\bibfnamefont {R.~T.}\ \bibnamefont
  {Scalettar}},\ and\ \bibinfo {author} {\bibfnamefont {G.~G.}\ \bibnamefont
  {Batrouni}},\ }\bibfield  {title} {\bibinfo {title} {Ring exchange and phase
  separation in the two-dimensional boson {H}ubbard model},\ }\href
  {https://doi.org/10.1103/PhysRevB.72.054524} {\bibfield  {journal} {\bibinfo
  {journal} {Phys. Rev. B}\ }\textbf {\bibinfo {volume} {72}},\ \bibinfo
  {pages} {054524} (\bibinfo {year} {2005})}\BibitemShut {NoStop}%
\bibitem [{Note3()}]{Note3}%
  \BibitemOpen
  \bibinfo {note} {We note that one can also consider $L=4m+2$ and $N=2m+1$
  with appropriate interchange of the roles of PBC and APBC. For example, it is
  then that the PBC ground state is non-degenerate, and lower in
  energy.}\BibitemShut {Stop}%
\end{thebibliography}%


\clearpage
\beginsupplement
\onecolumngrid
\vskip 10pt
\begingroup
  \centering
  \phantomsection
  \label{sec:supplemental-material}
  {\large\textbf{SUPPLEMENTAL MATERIAL}}\\[0.4em]
  {\large\textbf{Sign problem in world-line quantum Monte Carlo methods}}\\[0.6em]
  Rubem Mondaini and Richard Scalettar
\par
\endgroup
\vskip 8.5pt
\twocolumngrid
In the main text, we investigate how phase transitions (and the crossover toward a zero-temperature critical endpoint in one dimension) are encoded in the average fermion sign within world-line quantum Monte Carlo. This Supplemental Material provides the supporting analytical and numerical details. We further derive the exact finite-size average sign in one dimension, explain the thermodynamic-integration procedure used to reconstruct the free-energy difference in two dimensions, and show how the residual combination isolates its singular finite-size contribution. 


\section{World-line quantum Monte Carlo simulations}
\label{sec:sm-methods}
\begin{figure}[th!]
  \centering
  \includegraphics[width=\columnwidth]{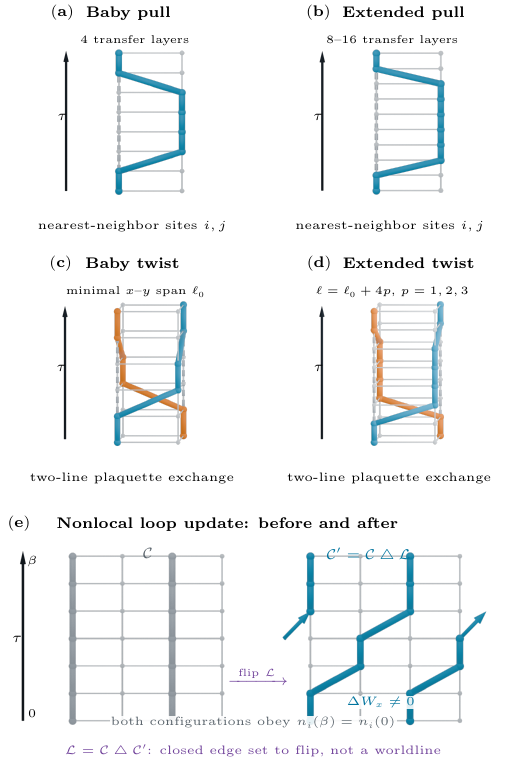}
  \caption{World-line updates used in the two-dimensional checkerboard implementation. Spheres denote the discrete space-time vertices. In panels (a)--(d), gray dashed paths show the original world lines and colored paths the proposed ones. (a) Baby pull: a world line is shifted to a nearest-neighbor site over one four-layer checkerboard period. (b) Extended pull: the same two-site contour spans two to four periods. (c) Baby twist: two world-line segments are exchanged across a spatial plaquette over the minimal separation between selected $x$- and $y$-bond layers. (d) Extended twist: the plaquette contour spans an additional one to three periods. These local moves conserve particle number and spatial winding. $(e)$ Nonlocal loop update. The auxiliary graph loop $\mathcal L$ specifies the edges whose occupations are toggled, $\mathcal C'=\mathcal C\mathbin{\triangle}\mathcal L$, and is not itself a world line. The update preserves temporal periodicity but may change spatial winding and permute world lines. The loop is constructed from compatible leg pairings at visited transfer vertices; proposals outside the fixed-particle-number sector are rejected, while allowed proposals receive a Metropolis correction to the interacting target weight.}
  \label{fig:wlqmc-updates}
\end{figure}
We provide here additional details of the world-line quantum Monte Carlo (WLQMC) calculations used in the main text.  All simulations are performed in the canonical ensemble at half filling.  Thus, for a lattice with $N_s$ sites, the total particle number is fixed to $N=N_s/2$. Periodic boundary conditions are imposed in space and in imaginary time.

\subsection{Discrete imaginary-time representation}

We employ a discrete imaginary-time representation, $ \beta=L_\tau\Delta\tau$, and decompose the kinetic Hamiltonian into sets of nonoverlapping bonds in the usual checkerboard construction~\cite{Scalettar1999, AssaadEvertz2008}. All results in the main text are obtained using $t\Delta\tau=0.05$. The corresponding Trotter error is of order $O(t \Delta\tau^2)$ and hence considerably less than one percent.

A world-line configuration specifies the particle occupation on every site $i$ and imaginary-time slice $\tau$, subject to the hard-core constraint $n_{i,\tau}=0,1$ and conservation of the total particle number.  For fermions, a configuration acquires a sign determined by the parity of the permutation of the world lines after propagation through the full imaginary-time
interval,
\begin{equation}
    {\cal S}({\cal C})=(-1)^{P({\cal C})}.
    \label{eq:sm-config-sign}
\end{equation}
If one removes this permutation sign while retaining the same local matrix elements, one maps the sampling problem onto the corresponding hard-core-boson system. This positive-weight bosonic ensemble is the reference ensemble used throughout the work.

\subsection{World-line updates}
\label{sec:sm-worldline-updates}

The configurations are sampled using five update families. The four local contour updates are illustrated in Fig.~\ref{fig:wlqmc-updates}(a)--(d): the baby and extended pulls rearrange one world line on a nearest-neighbor bond, whereas the baby and extended twists rearrange two world lines on a spatial plaquette -- these are borrowed from the local updates in Ref.~\cite{Rousseau2005}. Moreover, these local moves conserve both the total particle number and the spatial winding sector.

The nonlocal move, illustrated in Fig.~\ref{fig:wlqmc-updates}(e), is most
clearly described by regarding a configuration $\mathcal C$ as the set of
occupied space--imaginary-time edges.  At every visited transfer vertex, the
graph construction samples one of the two compatible pairings of its four
legs.  Following these pairings until the oriented starting leg is reached
produces a closed graph contour $\mathcal L$.  The proposal toggles the
occupation of every edge belonging to this contour,
\begin{equation}
  n'_e=
  \begin{cases}
    1-n_e, & e\in\mathcal L,\\
    n_e,   & e\notin\mathcal L,
  \end{cases}
  \qquad
  \mathcal C'=\mathcal C\mathbin{\triangle}\mathcal L.
  \label{eq:sm-graph-loop-update}
\end{equation}
Thus, $\mathcal L$ is not a particle world line.  It is the symmetric
difference between the old and proposed physical configurations, namely the
closed set of edges whose occupations change.  Accordingly,
Fig.~\ref{fig:wlqmc-updates}(e) displays the before-and-after physical
configurations $\mathcal C$ and $\mathcal C'$, with $\mathcal L$ appearing only
as the operation connecting them.  Compatibility of the local pairings
preserves particle conservation at every transfer vertex.  Proposals that
leave the fixed-$N$ sector are rejected, while the remaining proposals are
accepted with a Metropolis probability that corrects the noninteracting graph
proposal to the interacting target weight.

Both $\mathcal C$ and $\mathcal C'$ satisfy the imaginary-time boundary
condition $n_{i,L_\tau}=n_{i,0}$.  For indistinguishable fermions, an individual
labeled trajectory need close only up to a permutation; the parity of that
permutation gives the configuration sign in
Eq.~\eqref{eq:sm-config-sign}.  A closed graph contour can nevertheless wind
around a periodic spatial direction and thereby change $(W_x,W_y)$.  In one
dimension, for example, local updates starting from straight world lines
remain in the zero-winding sector, where the fermionic sign is positive.
Allowing winding-changing graph loops connects sectors with nontrivial
permutations and therefore samples both signs of the fermionic weight.

\subsection{Thermodynamic integration}

At sufficiently large system sizes or low temperatures, direct sampling of
the sign reaches its statistical floor.  We then reconstruct the same
free-energy difference from energy densities.  Defining
\begin{equation}
    \Delta f_L(\beta)
    =
    f_{{\rm F},L}(\beta)
    -
    f_{{\rm B},L}(\beta)
\end{equation}
and
\begin{equation}
    g_L(\beta)
    \equiv
    \beta\Delta f_L(\beta)
    =
    -\frac{1}{N_s}
    \ln\langle{\cal S}\rangle_{{\rm B},L},
    \label{eq:sm-g-method}
\end{equation}
one has
\begin{equation}
    \frac{d g_L}{d\beta}
    =
    e_{{\rm F},L}(\beta)
    -
    e_{{\rm B},L}(\beta).
    \label{eq:sm-g-derivative-method}
\end{equation}
At $\beta=0$, the canonical fermionic and hard-core-boson Hilbert spaces
have the same dimension and hence
\begin{equation}
    g_L(0)=0.
\end{equation}
It follows that
\begin{equation}
    \beta\Delta f_L(\beta)
    =
    \int_0^\beta d\beta'\,
    \left[
        e_{{\rm F},L}(\beta')
        -
        e_{{\rm B},L}(\beta')
    \right].
    \label{eq:sm-TI-method}
\end{equation}

Here $e_{{\rm F},L}$ is obtained from the canonical free-fermion calculation
above, while $e_{{\rm B},L}$ is measured in the sign-free hard-core-boson
ensemble.  The energy-density difference is evaluated on a discrete grid
of inverse temperatures and interpolated between neighboring simulated
points before performing the integration.  The corresponding diagnostic
curves are shown in
Fig.~\ref{fig:thermodynamic-integration-diagnostics-sm}.

Where the direct sign remains statistically resolved, we compare the
thermodynamic-integration result with the independent estimator
\begin{equation}
    \Delta f_L
    =
    -\frac{
        \ln\langle{\cal S}\rangle_{{\rm B},L}
    }{
        \beta N_s
    }.
    \label{eq:sm-direct-df-method}
\end{equation}
The agreement in this regime provides a direct check of the integration
procedure before it is extended into the regime in which
$\langle{\cal S}\rangle_{\rm B}$ can no longer be measured accurately.

\section{Exact average sign and finite-size scaling in one dimension}
\label{sec:sm-exact-sign-1d}

The flow of the one-dimensional sign crossover toward $T=0$ can be understood entirely from the finite-size spectrum, without Monte Carlo sampling. We take $L=4m$, so that half-filling corresponds to an even particle number, $N=L/2=2m$. The physical fermions obey periodic boundary conditions (PBC), whereas the Jordan-Wigner image of the periodic hard-core-boson reference system obeys antiperiodic boundary conditions (APBC)~\cite{Jordan1928,Lieb1961}. The average sign is therefore
\begin{equation}
    \langle{\cal S}\rangle_{\rm B}
    =
    \frac{{\cal Z}_{\rm F,PBC}(N,\beta)}
         {{\cal Z}_{\rm F,APBC}(N,\beta)}.
    \label{eq:sm-sign-ratio}
\end{equation}

The full temperature dependence follows efficiently from the canonical partition functions. Rather than enumerate the $\binom{L}{N}$ many-body configurations, we extract the coefficient of order $N$ from the corresponding grand-canonical polynomial,
\begin{equation}
    {\cal Z}_{\rm F,BC}(N,\beta)
    =
    [z^N]\prod_{k\in{\rm BC}}
    \left(1+z e^{-\beta\epsilon_k}\right),
    \quad
    \epsilon_k=-2t\cos k.
    \label{eq:sm-canonical-coefficient}
\end{equation}
where $[z^N]$ denotes the coefficient of $z^N$, and the allowed momenta are $k_n^{\rm PBC}=2\pi n/L$ and $k_n^{\rm APBC}=(2n+1)\pi/L$. This coefficient can be accumulated recursively. If $C_p^{(j)}$ is the coefficient of $z^p$ after the first $j$ single-particle levels have been included, then
\begin{equation}
    C_p^{(j)}
    =
    C_p^{(j-1)}
    +
    e^{-\beta\epsilon_j}C_{p-1}^{(j-1)},
    \qquad
    C_0^{(0)}=1,
    \label{eq:sm-canonical-recursion}
\end{equation}
with all other $C_p^{(0)}$ equal to zero. After the last level, ${\cal Z}_{\rm F,BC}=C_N^{(L)}$. Applying this recursion to the two spectra separately produces the complete finite-temperature curves in Fig.~\ref{fig:exact-sign-1d} of the main text.

The scale that governs the low-temperature crossover is already visible in the two ground-state sectors. For PBC, the $2m-1$ states with $|k|<\pi/2$ are occupied together with either one of the two zero-energy states at $k=\pm\pi/2$. This choice makes the PBC ground state twofold degenerate, with energy
\begin{equation}
    \begin{aligned}
        E_{0,\mathrm{PBC}}
        &=
        -2t\left[
        1+2\sum_{n=1}^{m-1}
        \cos\left(\frac{2\pi n}{L}\right)
        \right] \\
        &=
        -2t\cot\left(\frac{\pi}{L}\right).
    \end{aligned}
    \label{eq:sm-pbc-ground-state}
\end{equation}
For APBC, the $2m$ occupied states instead form pairs with momenta $\pm(2n+1)\pi/L$. The resulting ground state is nondegenerate, with energy
\begin{equation}
    \begin{aligned}
        E_{0,\mathrm{APBC}}
        &=
        -4t\sum_{n=0}^{m-1}
        \cos\left(\frac{(2n+1)\pi}{L}\right) \\
        &=
        -2t\csc\left(\frac{\pi}{L}\right).
    \end{aligned}
    \label{eq:sm-apbc-ground-state}
\end{equation}
The two expressions give the exact sector splitting
\begin{equation}
    \begin{aligned}
        \Delta E_L
        &=
        E_{0,\mathrm{PBC}}-E_{0,\mathrm{APBC}} \\
        &=
        2t\left[
        \csc\left(\frac{\pi}{L}\right)
        -
        \cot\left(\frac{\pi}{L}\right)
        \right] \\
        &=
        2t\tan\left(\frac{\pi}{2L}\right)
        =
        \frac{\pi v_{\rm F}}{2L}
        +O(L^{-3}),
    \end{aligned}
    \label{eq:sm-sector-splitting}
\end{equation}
where $v_{\rm F}=2t$ at half filling. Thus the APBC sector lies lower by an amount of order $1/L$. At sufficiently low temperature, the ratio of partition functions becomes a competition between this energy difference and the twofold PBC degeneracy,
\begin{equation}
    \langle{\cal S}\rangle_{\rm B}
    \simeq
    2e^{-\beta\Delta E_L},
    \label{eq:sm-low-temperature-sign}
\end{equation}
and the average sign vanishes as $T\rightarrow0$ for every finite ring\footnote{We note that one can also consider $L=4m+2$ and $N=2m+1$ with appropriate interchange of the roles of PBC and APBC.  For example, it is then that the PBC ground state is non-degenerate, and lower in energy.}.

Fixing the sign at any value ${\cal S}_\times$ defines a crossover temperature that simply tracks the sector splitting. Within the ground-state approximation,
\begin{equation}
    T_\times
    \simeq
    \frac{\Delta E_L}{\ln(2/{\cal S}_\times)},
    \qquad
    \frac{LT_\times}{v_{\rm F}}
    \simeq
    \frac{\pi}{2\ln(2/{\cal S}_\times)}.
    \label{eq:sm-crossover-scale}
\end{equation}
For ${\cal S}_\times=1/2$, as used in Fig.~\ref{fig:exact-sign-1d} of the main text, this estimate gives $LT_\times/v_{\rm F}\simeq\pi/(2\ln4)=1.1331$. The complete canonical partition functions yield $1.1397$. This small shift is expected: when $T\sim1/L$, the excitation energies scale in the same way, so excited states remain weakly populated even as $L$ grows. More importantly, both calculations give $T_\times\propto1/L$ and hence $T_\times\to0$ in the thermodynamic limit.

The collapse follows from the same low-energy scale. The half-filled chain has a gapless, $z=1$ spectrum~\cite{Haldane1981}, so its finite-size scale is $v_{\rm F}/L$. Comparing it with $T$ gives the variable $LT/v_{\rm F}$ observed in Fig.~\ref{fig:exact-sign-1d}(c). At fixed $L$, cooling selects the lower APBC sector and the sign vanishes; at fixed $T>0$, increasing $L$ removes the boundary-condition difference and the sign approaches unity.

Thus the extrapolation identifies a zero-temperature endpoint. Finite-range one-dimensional quantum lattices have no transition at $T>0$~\cite{Araki1969}; the hard-core-boson chain is critical only in its ground state, hence $T_{\rm KT}^{\rm(1D)}=0$.

\section{Thermodynamic integration in two dimensions - Thermal phase transition}
\label{sec:sm-thermodynamic-integration}

In the main text, we argue that one can bypass the statistical limitation on obtaining the average value of an exponentially small quantity, the average sign $\langle {\cal S}\rangle_{\rm B}$, via thermodynamic integration. We provide further details, including an explicit analysis of the free-energy difference for the case of the thermal phase transition analyzed there.

\begin{figure}[t]
    \centering
    \includegraphics[width=\columnwidth]
    {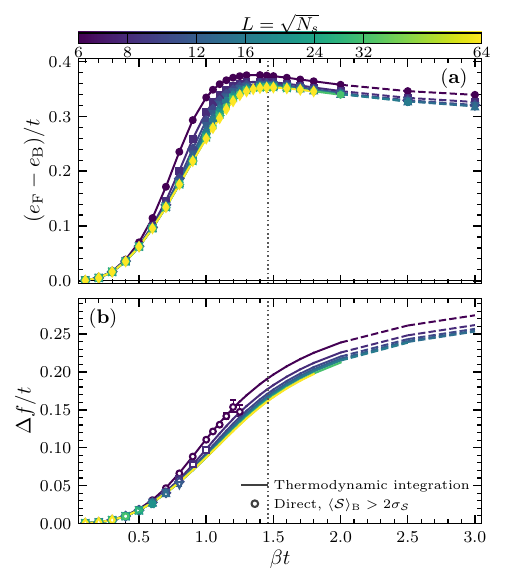}
    \caption{Thermodynamic-integration diagnostics for half-filled, noninteracting spinless fermions on the square lattice, using the hard-core-boson world-line reference at $t\Delta\tau=0.05$. (a) Energy-density difference $(e_{\rm F}-e_{\rm B})/t$; error bars show the propagated statistical uncertainty of the bosonic energy. (b) Free-energy-density difference $\Delta f=f_{\rm F}-f_{\rm B}$, reconstructed using $\beta\Delta f=\int_0^\beta d\beta'\,[e_{\rm F}(\beta')-e_{\rm B}(\beta')]$. Open symbols give $-\ln\langle{\cal S}\rangle_{\rm B}/(\beta N_s)$ only where the direct sign exceeds twice its standard error. Dashed segments span adjacent data points separated by $\Delta(\beta t)>0.35$; the vertical dotted line marks the fixed reference $\beta_{\rm BKT}t=1.460$. The fermionic energy uses the continuum-time canonical spectrum, while the bosonic data have finite $\Delta\tau$.}
    \label{fig:thermodynamic-integration-diagnostics-sm}
\end{figure}

Figure~\ref{fig:thermodynamic-integration-diagnostics-sm}(a) displays the integrand entering Eq.~\eqref{eq:integrated-sign} of the main text. The energy-density difference grows smoothly upon cooling, develops a broad maximum near the BKT temperature, and then decreases. The vertical dotted line denotes the independently determined value $\beta_{\rm BKT}t=1.460$ obtained from the finite-size scaling of the superfluid density (see Fig.~\ref{fig:sign-thermodynamic-integration} of the main text).

The resulting free-energy-density difference is shown in Fig.~\ref{fig:thermodynamic-integration-diagnostics-sm}(b). Wherever the direct sign satisfies $\langle{\cal S}\rangle_{\rm B}>2\sigma_{\cal S}$, the open symbols show the independently evaluated quantity
$-\ln\langle{\cal S}\rangle_{\rm B}/(\beta N_s)$ and provide a consistency check on the integration. Thermodynamic integration remains applicable after the direct estimator reaches its statistical resolution floor because it involves the difference of two intensive energy densities. It, of course, does not eliminate the sign problem: the factor $N_s$ in Eq.~\eqref{eq:integrated-sign} of the main text amplifies any residual integration error when the average sign is reconstructed. 

We note that the smoothness of the curves in Fig.~\ref{fig:thermodynamic-integration-diagnostics-sm} is consistent with the nature of the BKT singularity. On approaching the transition from the high-temperature side, with
\begin{equation}
    \tau
    =
    \frac{T-T_{\rm BKT}}{T_{\rm BKT}}>0\ ,
\end{equation}
the correlation length and the singular part of the bosonic free-energy density behave asymptotically as~\cite{Kosterlitz1974, Pelissetto2013}
\begin{align}
    \xi(\tau)
    &\sim
    \xi_0\exp\left(\frac{b}{\sqrt{\tau}}\right)\ ,\notag\\
    \quad
    f_{{\rm B},{\rm sing}}(\tau)
    &\sim
    A\xi^{-2}
    \sim
    A\exp\left(-\frac{2b}{\sqrt{\tau}}\right),
    \label{eq:sm-bkt-essential-singularity}
\end{align}
where $\xi_0$, $b$, and $A$ are nonuniversal. Because the free-fermion free energy is \emph{analytic} at this finite temperature, $\Delta f_{\rm sing} = -f_{{\rm B},{\rm sing}}.$ Nonetheless, displaying $f_{\rm B}$ instead of $\Delta f$ would therefore only reverse the singular contribution and modify the analytic background; it would not produce a visible cusp. And this is more evident if we recall that the essential singularity in Eq.~\eqref{eq:sm-bkt-essential-singularity} is exponentially weak, and its finite-order temperature derivatives vanish as $\tau\rightarrow0^+$.

Consequently, Fig.~\ref{fig:thermodynamic-integration-diagnostics-sm} is merely a validation of the thermodynamic integration. Indeed, the critical finite-size contribution is exposed by combining the same free-energy data at sizes $L$ and $2L$, as we do in the main text:
\begin{align}
    R_L^{({\cal S})}(\beta)
    &=
    -\beta L^2
    \left[
        \Delta f_L(\beta)-\Delta f_{2L}(\beta)
    \right]\notag \\
    &=
    \ln\langle{\cal S}\rangle_{{\rm B},L}
    -
    \frac{1}{4}
    \ln\langle{\cal S}\rangle_{{\rm B},2L}\ .
    \label{eq:sm-residual-from-free-energy}
\end{align}
This difference cancels the leading regular bulk contribution and isolates the order-unity finite-size term whose logarithmic BKT scaling is examined in Fig.~\ref{fig:kt_residual} of the main text.

\end{document}